\documentclass[journal=jacsat,manuscript=article]{achemso}
\usepackage{multirow}
\usepackage{xcolor}
\usepackage{xr}
\usepackage[version=3]{mhchem} % Formula subscripts using \ce{}

\author{Yao Wei}
\affiliation{Theory and Simulation of Condensed Matter
(TSCM), King's College London, Strand, London WC2R 2LS, United Kingdom}
\author{Alejandro Santana-Bonilla}
\affiliation{Theory and Simulation of Condensed Matter
(TSCM), King's College London, Strand, London WC2R 2LS, United Kingdom}
\author{Lev Kantorovich}
\affiliation{Theory and Simulation of Condensed Matter
(TSCM), King's College London, Strand, London WC2R 2LS, United Kingdom}
\email{lev.kantorovitch@kcl.ac.uk}

\title{Global Optimization of Molybdenum Subnanoclusters on Graphene: a Consistent Approach Towards Catalytic Applications}

\abbreviations{IR,NMR,UV}
\begin{document}


\begin{abstract}
  The development of novel sub-nanometer clusters (SNCs) catalysts with
superior catalytic performance depends on the precise control of clusters'
atomistic sizes, shapes, and accurate deposition onto surfaces. 
The intrinsic
complexity of the adsorption process complicates the ability to achieve
an atomistic understanding of the most relevant structure-reactivity
relationships hampering the rational design of novel catalytic materials.
In most cases, existing computational approaches rely on just a few
structures to draw conclusions on clusters' reactivity thereby neglecting
the complexity of the existing energy landscapes thus leading to insufficient
sampling and, most likely, unreliable predictions. Moreover, modelling
of the actual experimental procedure that is responsible for the deposition
of SNCs on surfaces is often not done even though in some cases this procedure may enhance the significance of certain (e.g., metastable)
adsorption geometries. This study proposes a novel systematic approach that utilizes global search techniques, specifically, the particle swarm optimization (PSO) method, in conjunction with \textit{ab-initio} calculations, to simulate {\it all stages} in the beam experiments, from predicting the most relevant SNCs structures in the beam and on a surface, to their reactivity. To illustrate the main steps of our approach, we consider the deposition of Molybdenum SNC of 6 Mo atoms on a free-standing graphene surface, as well as their catalytic properties with respect to the CO molecule dissociation reaction. Even though our calculations are not exhaustive and serve only to produce an illustration of the method, they are still able to provide insight into the complicated energy landscape of Mo SNCs on graphene demonstrating the catalytic activity of Mo SNCs and the importance of performing statistical sampling of available configurations. This study 
establishes a reliable procedure for performing theoretical rational
design predictions.


\end{abstract}

\newpage 

\section{Introduction}
Transition-metal sub-nanometer clusters (SNCs) have emerged as promising
novel catalytic materials, demonstrating superior performance compared
to their metallic and transition metal nanoparticle counterparts  \cite{James2015Energy}.
This remarkable enhancement arises from their high surface-to-volume
ratio, which facilitates a unique catalytic behaviour due to a combination
of low atomic coordination number and electronic accessibility to
transition-metal atoms to reactants \cite{campelo2009sustainable}. Advancements in synthesis
and characterization techniques have enabled the production of atomically
precise transition-metal SNCs, with the ultimate experimental objective
being the development of scalable and cost-effective methods to synthesize
these materials with tailored properties \cite{C2CS15325D}.
Therefore, developing a rational design strategy is paramount to effectively
deploy these materials in a diverse range of applications, such as energy
storage, photocatalysis, and biomedicine \cite{Oxygen_Electroreduction,muhammed2009bright,santiago2010one}.

Despite this experimental progress, transition-metal SNCs are likely
to sinter and lose surface area when used as catalysts \cite{fernandez2018sub}.
To prevent this, supported solid-state materials (such as carbon or
metal oxide surfaces) are used to stabilize transition-metal SNCs \cite{zhang2022supported,pacchioni2013electronic}
. Among many experimental
techniques, the beam deposition methods \cite{wegner2006cluster}
have been successfully tested for depositing on different substrates
atomically precise transition metal (TM) clusters with a pre-defined
number of atoms. In these methods, an induced positive charge of the
clusters enables one to direct them precisely onto the surface, while
their low kinetic energies prevent the clusters from decomposing when
landing on it (the so-called ``soft landing''). Moreover, it is
assumed that upon contact the clusters instantaneously get the electronic
charge from the surface and end up adsorbed in the neutral state.

Due to the complexity of the atomic-level process described, developing
a theoretical understanding of the structure-reactivity relationship
for the clusters on the surface is crucial \cite{halder2018perspective}. From an atomistic perspective,
exploring the accessible thermodynamic states of the clusters on the
surface is challenging due to the huge variety of possible clusters
and numerous degrees of freedom that must be considered. Fortunately,
the atomic-scale control offered in the beam experiments \cite{Loi-Baraldi-2022}
(whereby the number of atoms in the clusters is known) reduces enormously
the amount of theoretical work required to forecast clusters' adsorption
geometries on the surface of interest, calculate their electronic
structure, including their magnetic states, and consequently predict
their reactivity (catalytic activity) towards specific chemical reactions.
%of interest. 
Note that, from the theoretical point of view, the problem
would otherwise be unmanageable due to the extreme complexity of the
energy landscape and hence the necessity of considering an enormous
number of possible adsorption systems with a wide range of adsorption
energies.

Even due to the mentioned significant reduction of possibilities, 
%to consider because of the known cluster sizes, 
one still needs to find the most probable
adsorption geometries to be realized in the beam experiments, some of which
%of such geometries 
(or all) could be metastable. Straightforward geometry
optimization strategies relying on guessing initial geometries are still abundant in the literature, for recent examples see, e.g., Refs. \cite{Huang-2023,Ghosh-2024,Zaman-2020,Jin-2022,hao2018coverage,ghosh2013fluxionality,zhang2024promoting}. However,  these naive approaches are
most likely inapplicable here. Modern computational procedures, such
as meta-heuristic algorithms including genetic algorithms \cite{chen2017improved,deaven1995molecular,davis2015birmingham,kanters2014cluster,alexandrova2010h,Liu-2020,Heard-2014}, random sampling \cite{pickard2006high,pickard2009structures,pickard2011ab},  data mining \cite{fischer2006predicting}, simulated annealing \cite{pannetier1990prediction,wang2009structural}, basin hopping, \cite{Wales-book} and Particle Swarm Optimization (PSO) \cite{wang2012calypso}, can provide a more effective search process by employing efficient exploration techniques of the configurational space associated with a specific optimization problem. Mentioned above and some other global optimization techniques are reviewed in Refs. \cite{Oganov-Ed-book-2010,zhai2024benchmarking,zandkarimi2019surface}. Moreover, an additional advantage of most of these methods is that they also offer a range of metastable geometries with energies near the global minimum,  which may even be more reactive \cite{zandkarimi2019surface}.

In this work, we employ global minimum search methods, such as the
\emph{ab initio} random structure searching (AIRSS) \cite{pickard2006high,pickard2011ab}
and PSO algorithm \cite{wang2018particle,shi2004particle}
to investigate the energy landscape of a series of Molybdenum SNCs
adsorbed onto pristine free-standing graphene in the beam experiments.
 Note that the PSO algorithm was highly rated in a recent study \cite{zhai2024benchmarking} for its stable performance  compared to other methods.
The adsorption of transition-metal SNCs onto an inorganic support
profoundly influences their properties, including geometry, electronic
structure, and charge state, ultimately dictating the catalytic performance
of the system. We have chosen Molybdenum as it is well known for its
catalytic properties \cite{H-evolution-reaction-Mo-catalysts-Rare-Met-2020,H-evolution-reaction-Mo-catalysts-Systainability-2023,Mo-watr-air-pipputants-EcoMat-2021}.
Using a unified approach based on stochastic methods, we study the
entire experimental procedure, starting from the clusters in the beam
and their adsorption on the surface, and then moving on to investigate
their reactivity with respect to the dissociation of a CO molecule.
The results of these studies indicate that the predicted Molybdenum
SNCs do indeed exhibit catalytic activity.

Literature on combined Molybdenum and CO molecule systems is rather scarce. There have been some experimental and computational studies on the interaction of the CO with Mo(100), Mo(110) and Mo(112) surfaces, in which dissociation of the CO has also been discussed \cite{Jiel-2003,Yakovkin-2009,Yang-2009,RAAEN201517,Tian_2017,Gleichweit-2017,CO-diss-on-Mo110}. We are also aware of a single study of the chemical reactivity of Mo$_n$ clusters (with $n\le14$) towards binding the CO molecule \cite{cox1988co}, in which, however, dissociation of the latter was not considered. In the computational study \cite{addicoat2008associative} CO dissociation on the Mo$_3$ gas phase cluster was investigated where a few eV energy barrier was obtained. To the best of our knowledge, there have been no studies of reactivity of Mo clusters adsorbed on graphene or graphite surfaces towards the CO dissociation reaction, neither experimental nor computational.

If each particular step of the experimental procedure has been simulated and reported previously, e.g., gas-phase clusters or clusters' adsorption on a surface, we are not aware of a consistent and systematic global optimization approach applied  throughout all steps of the whole procedure.
%, from clusters' formation in the beam to their surface adsorption and reactivity. 
To the best of our knowledge, this is the first study of that kind that simulates computationally   the entire experimental protocol, from the formation of the clusters in the beam and their adsorption on a surface to a chemical reaction catalysed by them.

We stress that the main purpose of this work is to emphasize the importance of thoroughly exploring the potential energy landscape in investigating
every step of the experimental procedure, and that, due to overall complexity of the problem, stochastic methods must
be used at each stage to properly sample the corresponding configurational phase space.
%, from the clusters' formation,
%their adsorption on a surface and subsequently their role in the chemical
%reaction of interest. 
Following this paradigm in full would require
substantial computational resources. Therefore, in this work, as the first step towards the full approach, we have just tried to \emph{illustrate}
the method by selecting a simple system (Mo clusters of 6 atoms and
the free-standing graphene as a surface) and follow at each step
just a subset of available opportunities. We have found this procedure
sufficient to understand the multitude of various possibilities one
has to tackle when studying these kinds of problems, and appreciate
the necessity of thoroughly exploring the potential energy surface.
A fully comprehensive study is left for future work.

\section{Methods}

\subsection{General computational strategy and computational details\label{subsec:General-computational-strategy}}

\begin{figure*}[h]
\centering{}\centering \includegraphics[width=1\textwidth]{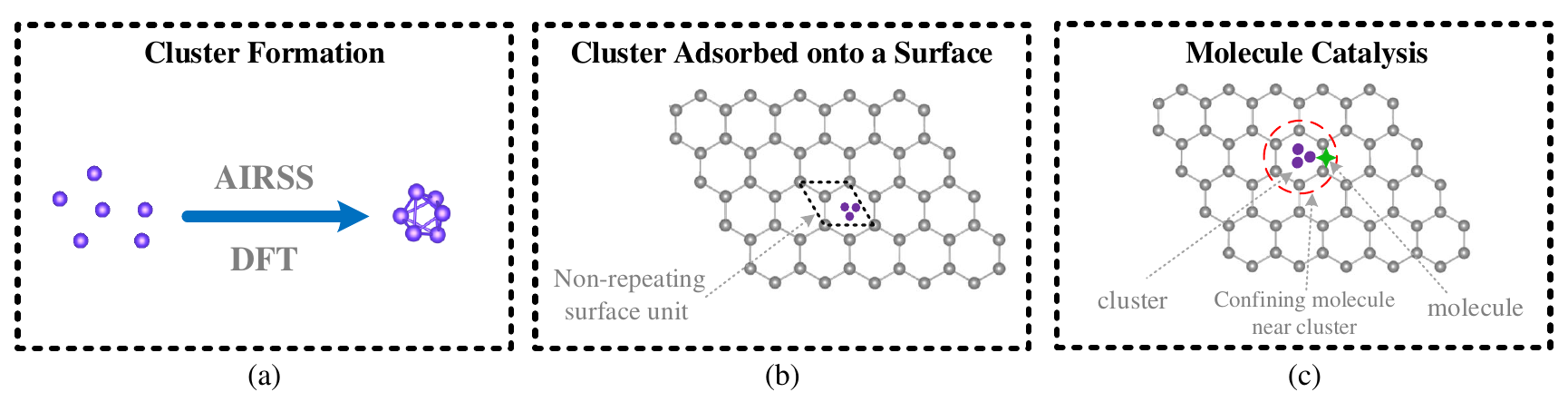}
\caption{Schematic Representation of the simulation approach followed in this
study. (a) Positively charged clusters in the gas phase.
 (b) Clusters adsorbed onto a surface. (c) Consideration of the catalysis at and around the cluster.}
\label{diagram1}
\end{figure*}

We propose the following \emph{general} computational strategy that
can be split into several interconnected stages:
\begin{enumerate}
\item In the beam experiments \cite{Loi-Baraldi-2022,heinrich2000cluster} monoatomic clusters
are ejected by an atomic gun and ionized. Carrying a single positive
charge, it is then possible to affect the direction they
move in, select the clusters of a particular mass and, therefore, of the
desired number of atoms, Fig. \ref{diagram1}(a). This means that one
can experimentally select the monoatomic clusters of a pre-defined number
of atoms in the beam and direct them onto the surface of interest
with great precision. Moreover, because of the clusters' charge, it
is possible to control their kinetic energy, which enables one to
land the clusters ``softly'' on the surface. From theoretical point of view, this means that one can use
their equilibrium gas-phase geometry as the initial structure, prior to geometry relaxation
due to the interaction with the surface. 
The soft landing of Au clusters on the TiO$_2$ surface was, e.g.,  modelled using molecular dynamics simulations in Ref. \cite{li2015structure}.
It is well known that the
final geometry in the geometry relaxation process in many cases is influenced by
the initial geometry, i.e. instead of the global energy minimum, metastable
structures would normally be established. Hence, to find
clusters' geometries on the surface, one needs first to find the lowest
energy structures of positively charged clusters in the gas phase.
Once these are known, we would be able to consider mostly the cluster(s)
of lowest energy(ies) in placing them on the surface, prior to geometry
relaxation. So, at the first stage of our computational approach,
we are interested in investigating possibly all lowest energy clusters
of a single positive charge in the gas phase.

\begin{figure*}[h]
\centering{}\centering \includegraphics[width=1\textwidth]{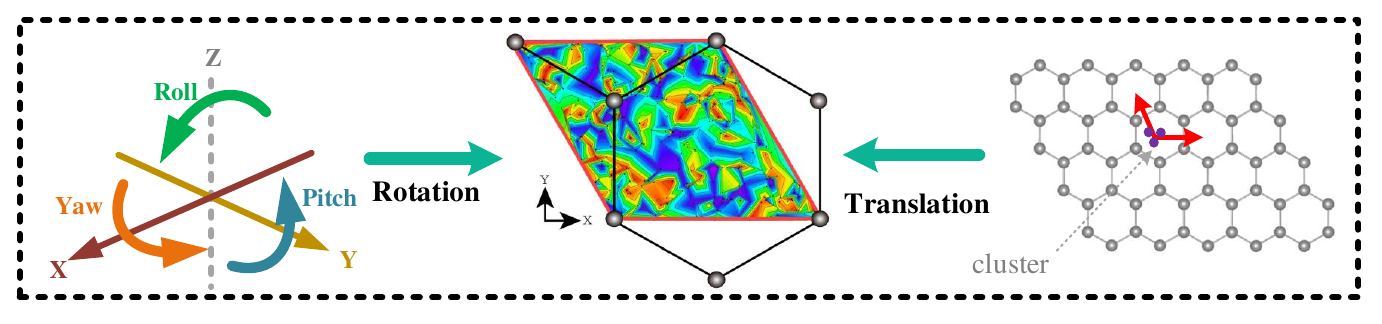}
\caption{Schematic representation of a 5D multi-objective PSO framework for surface-cluster adsorption. This scheme considers 3 rotations and 2 translations, yielding projected potential energy surface as a function of cluster exploration.}

%\caption{Schematic Representation of multi-degree-of-freedom PSO algorithm for surface cluster absorption prediction. 
%}
\label{diagram2}
\end{figure*}

\item At the next stage, we intend to put these clusters on the surface
of interest. However, this can be done in many ways as clusters would
explore the free-energy landscape generated by the surface, seeking
the most stable configuration for the attachment, as illustrated in
Fig. \ref{diagram2}. However, the most appropriate way of choosing
the clusters' adsorption sites is instigated by the experimental procedure
itself. Indeed, the clusters in the beam may land on the surface at
different lateral positions (via a uniform distribution in a uniform
beam) and at different spatial orientations (again, with equal probability).
Therefore, we have designed a procedure (to be detailed below) whereby
the lowest gas-phase energy clusters are placed on the surface at
random, at various lateral positions (within the periodic cell) and
various orientations in space. There are five variables here to
vary: two lateral positions and their angles that uniquely specify
the clusters' orientation. It is assumed that upon landing the clusters
will almost immediately (at electronic timescales) get neutralized
on the surface via an electron transfer from it \cite{popok2011cluster,johnson2016soft};
hence, we can proceed with the subsequent geometry relaxation of the
clusters on the surface considering the whole system is neutral.
After accomplishing this stage, we shall have the lowest energy structures
of the clusters on the surface that are attainable in this type of
experiment. It is essential to realize two points here: (i) most likely,
there will be many geometries of the clusters with energies close to
the lowest energy configuration, and hence all of them need to be taken
into consideration while studying applications, e.g., in catalysis, and
(ii) the obtained geometries of the clusters with $n$ atoms will
most likely \emph{not} correspond to the global energy minimum of
the surface with $n$ \emph{individual }atoms, as the obtained final
geometries will be influenced by the clusters' initial geometries
in the beam prior to their contact with the surface. It is believed
that the obtained geometries of the clusters on the surface would
correspond to the majority of such clusters in the considered experimental
procedure. Hence, finding the lowest energy structures of the clusters
in the gas phase and then exploring their adsorption comprises two
interlinked essential parts of our computational strategy that is
meant to model this particular beam experiment.
\item At the third stage we shall study the role of these clusters in catalyzing
a dissociation chemical reaction. To this end, we need first to find
all lowest energy adsorption sites of the molecule of interest at
or near one of the clusters found, to be considered as the initial
geometries of the reaction (reactants), Fig. \ref{diagram1} (c). Next,
in a similar fashion, all possible lowest energy structures of the
products of the reaction at or near the cluster are to be determined
using a similar procedure, to be considered as the final geometries
after the reaction. Then, all the initial and final structures need
to be connected by minimum energy path calculations to determine the
corresponding energy barriers and hence the transition rates. These
calculations are to be repeated for all lowest energy clusters on
the surface found.
\item The paths corresponding to the lowest energy barriers, in conjunction
with the lowest energy clusters (that determine their abundance on
the surface) are to be considered as \emph{defining }the chemical
reaction of interest. Hence, at the last stage, we consider kinetics
of the dissociation process in which all clusters participate and
all obtained dissociation paths are included explicitly. The initial
concentrations (or populations) of the various clusters prior to the
reaction (upon deposition) are determined by the canonical statistical
distribution based on their formation energies, and the evolution
in time of the products - by the obtained transition rates based on
the calculated energy barriers.  Since it is assumed that during the course of the reaction diffusion of reactants and their possible transformations are unlikely 
(which as well may not be true),
the rate equation approach \cite{seki2003fractional,hellander2015reaction,Rate-eq-KMC-Tetlow}
will be employed to study the time dynamics of the dissociation reaction.
If the surface diffusion  and structural fluxionality effects  \cite{zandkarimi2019surface,ghosh2013fluxionality,poths2024thermodynamic}
 are thought to play an essential role, methods
such as kinetic Monte Carlo (kMC) \cite{Fichthorn-Weinberg-kMC,Nitzan-2014}
will have to be employed (see, also \cite{Rate-eq-KMC-Tetlow,kMC-Tetlow,poths2024thermodynamic}).
\end{enumerate}
To accomplish each stage of the above procedure, we shall employ stochastic
approaches that are the most suitable in each case as detailed in
the next subsection.

\subsection*{Computational methods}

To obtain the total energies of the systems, we have used \emph{ab initio}
computational method based on the Density Functional Theory (DFT)
approach. All DFT calculations have been performed using the VASP
(Vienna Ab initio Simulation Package) code \cite{kresse1996efficient,vasp2kresse1996efficiency}
using the Perdew, Burke, and Ernzerhof (PBE) generalized gradient
approximation (GGA) functional \cite{perdew1996generalized}. Additionally,
the van der Waals (vdW) dispersion correction was introduced by employing
Grimme’s D3 method \cite{grimme2011effect}. Electron-ion interactions
were described using the projector-augmented wave (PAW) method \cite{civalleri2008b3lyp}
with electronic configurations $2s^{2}2p^{2}$ and $4s^{2}4p^{6}4d^{5}5s^{1}$
for C and Mo atoms, respectively. The Mo configuration chosen was
found necessary in our previous study of gas-phase Mo clusters \cite{wei2023comprehensive}.
The calculations were performed  using several values of the number of unpaired electrons in the system to account for  the possibility of different spin states. 
Due to the large sizes of the periodic cells used (see below), the $\Gamma$-point has been found sufficient to sample the Brillouin zone in all our calculations. A
plane-wave kinetic-energy cutoff of 400 eV was used, and each relaxation
the calculation was considered finished when the energy threshold of 10$^{-5}$
eV was reached, while the atomic forces were not larger than 0.01
eV/\AA. To perform DFT simulations on charged Mo clusters,
we have used a standard procedure implemented in VASP in which a neutralizing
background is implemented and a Coulomb correction is applied to weaken
the inevitable dependence on the cell size. When performing adsorption
calculations, we chose a simulation cell containing 72 carbon atoms
that is obtained by expanding the primitive cell of graphene (2 atoms
in the primitive cell) along two primitive lattice vectors by 5 and
6 times, respectively. To prevent the translation of the whole system
during the relaxation, a few atoms of graphene that are remote to
the cluster were kept fixed. Spin polarization is considered
in our calculations, employing an initial magnetic moment of 1.0 Bohr magnetons ($\mu_{\mathrm{B}}$) for each Mo atom.

To compute the minimum energy path in the CO dissociation reaction
between the initial position of the CO molecule near or at a Mo cluster
and the final positions of the individual C and O atoms after the
reaction, we used the Nudged Elastic Band (NEB) method \cite{henkelman2000improved}
as implemented in the CINEB+VASP code \cite{henkelman2000climbing}.
 An appropriate number of images (as indicated  below) was used in all NEB simulations and each calculation
was considered converged when the energy and forces were better than
10$^{-5}$ eV and 30 meV/\AA, respectively.

\subsection*{Algorithms' implementation}

At the first stage, to simulate the process of cluster formation in
the gas phase, we employed the AIRSS (Ab initio Random Structure Searching)
random structure search package \cite{pickard2011ab} to identify
the lowest energy gas-phase structures. AIRSS achieves this by considering,
at random, various physical features, such as atomic distances and
symmetry. The software systematically explores the configurational
space to identify promising atomic structures and has been successfully
applied by some of us previously to study neutral Molybdenum SNCs
\cite{wei2023comprehensive}. Using this approach we were not only
able to find Mo clusters proposed previously, but also to predict
many new ones, of very different geometry and electronic and magnetic
structure. The AIRSS approach is well suited for considering low-energy
structures formed by a set of \emph{individual atoms} as it enables
one to construct many  possible compositions.

To navigate the intricate configurational landscape created by clusters
on the surface, AIRSS is no longer appropriate as we need to consider
at random only the lateral position of the whole cluster of known
geometry and its orientation, 5 degrees of freedom in total
upon its adsorption on the surface. Hence, we have chosen the Particle Swarm Optimization (PSO) technique as a promising approach to effectively
pursue the vast array of possibilities arising from this process.
PSO is a population-based stochastic optimization method inspired
by the social interactions observed in bird flocks and fish schools.
In PSO, each particle represents a potential solution to the optimization
problem, with its position corresponding to a candidate solution and
its velocity governing its movement through the configurational space.
PSO has gained widespread use across various domains, including function
optimization, parameter tuning, data clustering, and crystal structure
prediction \cite{wang2012calypso}. Its simplicity, computational
efficiency, and ability to tackle multimodal and non-linear problems
make it a popular choice for optimization tasks. We stress that it
is not guaranteed that the best structure found by PSO is the true
global energy minimum; however, with each generation (iteration),
as the configurational space is increasingly better explored, the
best structures obtained become lower and lower in their energies,
thereby (at least) consistently approaching the global energy minimum.
Note that the method offers a possibility of keeping lower energy structures that are found during the course of the optimization as well, thereby providing, at no additional cost, a sample of many energetically favourable structures. 

In the PSO framework, each potential solution (in our case, this comprises
a 5-fold vector $\mathbf{X}$ of the lateral position and orientation
of a cluster) is represented by a particle, and a collection of these
particles at every given iteration forms a generation. These particles
navigate the search space, influenced by their own best-known positions
and the best positions of their neighbors. 
%The algorithm's steps are
%illustrated in Fig. \ref{diagram2}. 
Initially, a swarm of particles
with random positions and velocities is established within the search
space. Subsequently, the fitness of each particle is evaluated, reflecting
the quality of the solution it represents. In our case the fitness
is determined by the DFT energy of the structure obtained {\em after} geometry
relaxation from the initial geometry given by the vector $\mathbf{X}$. This was found essential as otherwise (if the PSO was run without geometry optimisation) some of the structures that upon optimisation have low energies may be rejected as they appear at the end of the list.  
In performing the geometry relaxation, the vertical position of each
cluster prior to the relaxation was in all cases chosen as 3.0 $\textrm{Å}$
between the graphene and the lowest cluster atom. The rationale for choosing this particular initial vertical position is to ensure that, on the one hand, the cluster will not crash onto the surface, and, on another, it is not too far from it either so that there is a numerically significant force to pull it down. The personal best
position $\mathbf{X}_{best}$ for each particle is updated, representing
the particle's most promising position explored so far. Next, the
particle with the best $\mathbf{X}_{best}$ across the whole swarm
of particles of the current generation is identified, representing
the current best solution found by the swarm. This particle's position
is termed the global best position $\mathbf{X}_{g-best}$. For a new
generation $t+1$, the ``velocity'' of $i$-th particle is updated
based on its current position $\mathbf{X}_{i}(t)$ in the current
generation $t$ as well as the values of \textit{$\mathbf{X}_{best}$}
and \textit{$\mathbf{X}_{g-best}$} following the formula:
\begin{equation}
\mathbf{V}_{i}(t+1)=w\mathbf{V}_{i}(t)+c_{1}r_{1}\left[\mathbf{X}_{best}(t)-\mathbf{X}_{i}(t)\right]+c_{2}r_{2}\left[\mathbf{X}_{g-best}(t)-\mathbf{X}_{i}(t)\right],
\end{equation}
where $\omega$ represents the inertia weight, while $r_{1}$ and
$r_{2}$ are uniformly distributed random numbers $U(0,1)$. The constant
$c_{1}$ determines the influence of a particle's own experience on
its movement (Personal Learning Rate), whereas $c_{2}$ determines
the influence of the swarm's collective knowledge (Social Learning
Rate). Finally, the position of the particle is updated via:

\begin{equation}
\mathbf{X}_{i}(t+1)=\mathbf{X}_{i}(t)+\mathbf{V}_{i}(t+1)\,,
\end{equation}
producing the particle's new location in the search space (a new generation).
For this work, we have used 0.1, 0.16, and 0.84 for $\omega$, $c_{1}$,
and $c_{2}$, respectively. 
These values for the parameters were determined through a series of benchmark tests aimed at optimizing the balance between exploration and exploitation within our specific model context.
We used a swarm of 20 particles (i.e different
initial cluster configurations). As was mentioned before, at each
step the position $\mathbf{X}_{i}(t)$ of the particle is used to
work out the initial geometry of the cluster on the surface prior
to the DFT geometry optimization (see the Supporting Information file)
in which both the graphene surface and the cluster atoms were allowed
to relax (apart from a few fixed graphene atoms, as mentioned earlier);
upon relaxation, the final DFT energy represents the fitness of the
particle in the swarm that determines its position in it and hence
which of the structures in the swarm is the best.

As was explained above, during the search for the optimal adsorption
configuration, only the lateral position, given by the centre of mass, and the orientation of the clusters, given by three rotational angles
(two translational coordinates and three angles), are employed
in the PSO algorithm (see the Supporting Information file for details). The sensitive nature of these five control parameters
during the cluster displacement and rotation makes it challenging
to avoid unfavourable local minima in the adsorption site search. To
mitigate this issue, approximately 40\% of the least favourable structures
were discarded in each generation, and their corresponding structural
parameters were randomly regenerated (so that each generation always
contained exactly 20 structures). The convergence of the generations
in finding the adsorption site with the lowest overall energy depends
on the complexity of the system; in current simulations, if the energy
difference of the best ten systems in the simulation is smaller than
0.01 eV, we considered it as converged. Depending on the system, between
15 to 40 generations were required to reach this convergence criterion.

A modified version of the PSO method has also been used in predicting
positions of the CO molecule prior to the dissociation reaction, as
well as in predicting the position of the O atom after the reaction.
To simplify these latter calculations, we have considered a single
position of the C atom on the cluster only. Of course, a comprehensive
approach, as was outlined in the previous subsection, would go through
all possible positions of both C and O atoms as the final states of
the reaction. That would undoubtedly increase the cost of these simulations,
let alone the number of possibilities to consider in rather expensive
NEB simulations that will follow. As our main aim in this work is
mainly to highlight the complexity of the problem at hand and to indicate
the appropriate workflow to tackle it, we find our limited approach
outlined above to be sufficient here for our purposes.

As the number of DFT geometry relaxations during the course of the
PSO search is counted into thousands and the size of the systems was
much larger than for the gas-phase clusters within the AIRSS approach,
larger tolerances of $5\times10^{-5}$ eV (energy) and $50$ meV/\AA
(forces) were used in all PSO simulations.

When using both AIRSS and PSO methods, hundreds of structures were
generated that span a wide range of energies (see below). We found
that in each case there are the significant number of structures generated
with their energies very close to the best energy we find. A detailed
analysis showed that many of the structures we find are essentially
identical, hence, a necessity arose when using these stochastic approaches to 
automatically identify the identical structures  and then 
discard them. Our method is close in spirit to the one considered in Ref. \cite{zhai2016ensemble}.

When performing AIRSS simulations on isolated clusters,
we identified identical geometries by considering their symmetries.
In PSO simulations for clusters on graphene, we encountered a complete
lack of symmetry in the relaxed clusters. As a result, the identification
of distinct adsorbed configurations became challenging through conventional
symmetry analysis. To address this issue, we adopted an alternative
approach by examining the spatial distribution of atoms within the
cluster relative to its centre of mass. Subsequently, we first organized
the configurations based on their energy in the ascending order. If the
spatial distributions were found to be closely aligned, indicating
a similarity, we classified them as duplicates and excluded from further
consideration. This methodology allowed us to discern and catalogue
identical geometries within the PSO simulations.

\section{Results}

\subsection{Lowest energy Mo clusters\label{subsec:Lowest-energy-Mo6}}

In our recent work \cite{wei2023comprehensive}, we conducted an
analysis of \emph{neutral} Mo clusters comprising 3 to 10 atoms. That
method was extended here for the investigation of Mo clusters
with a single positive charge. To illustrate our approach, only clusters
Mo$_{6}$ containing 6 Mo atoms are considered.

\begin{figure*}[htbp]
\centering
\includegraphics[width=1\textwidth]{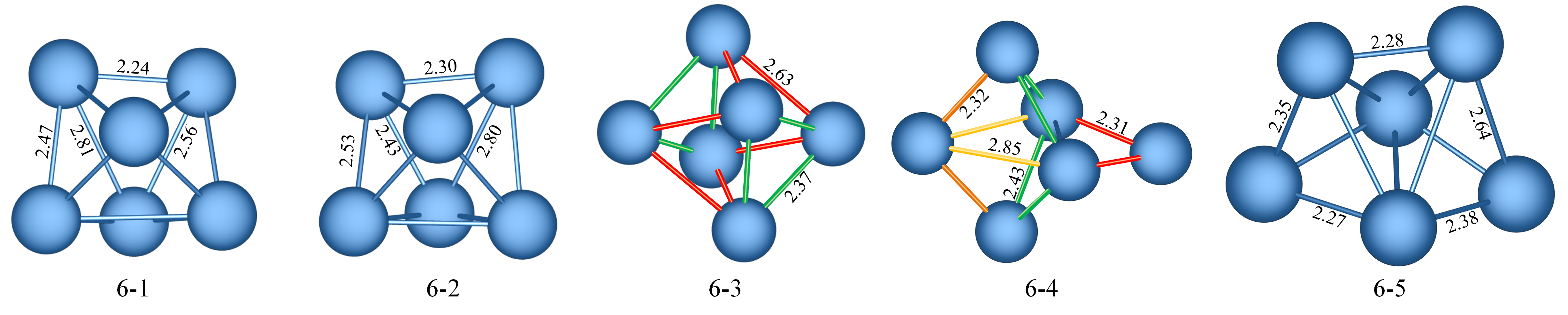}
\caption{Lowest energy configurations of charged 6-atom Mo clusters. Mo-Mo bonds of the same length in clusters 6-3 and 6-4 that possess symmetry are shown in the same colour for convenience.}
\label{6_atoms_charged_figure}
\end{figure*}

Using the AIRSS method, we found more than 70 stable clusters with
multiplicities equal to 2, 4 and 6. The total energies of the clusters
span a huge energy range of over 10 eV from -43.24 eV to -32.95 eV.
It is noted that all geometries within this energy range exhibit stability.
However, within the energy range of 1.1 eV from the lowest energy
cluster, only five clusters were found. These five lowest energy clusters
are depicted in Fig. \ref{6_atoms_charged_figure} and their properties
are shown in Table \ref{6_atoms_charged_table}.

\begin{table*}[htbp]
\centering
\caption{Energies and point group symmetries of the charged Mo$_{6}^{+}$ clusters.}
\label{6_atoms_charged_table}
\begin{tabular}{c|ccccc}
\hline 
 & 6-1 & 6-2 & 6-3 & 6-4 & 6-5\tabularnewline
\hline 
Total energy (eV) & -43.24 & -42.61 & -42.44 & -42.30 & -42.15\tabularnewline
\hline 
Multiplicity & 2 & 4 & 2 & 2 & 2\tabularnewline
\hline 
Symmetry & $C_{1}$ & $C_{1}$ & $D_{3d}$ & $C_{s}$ & $C_{1}$\tabularnewline
\hline 
\end{tabular}
\end{table*}

When contrasting these results with the outcomes observed for neutral
Mo$_{6}$ clusters, we notice that the introduction of a single positive
charge induces a reduction in the symmetry of the cluster as the lowest
energy configuration exhibits $C_{1}$ symmetry, whereas the neutral
Mo$_{6}$ lowest energy cluster shows $C_{2v}$ symmetry. It is noteworthy that the cluster with the lowest energy is 0.6 eV lower in energy compared to the cluster with the second lowest energy. Note that the lowest energy cluster with multiplicity 6 (not shown) is nearly 1.2 eV higher in energy than the lowest energy cluster. It is clear that, because of the considerable energy difference, for this particular Mo$_{6}$ system, it would be sufficient to proceed only with the single cluster of the lowest energy when placing the cluster on graphene and performing CO dissociation
simulations, as described in the forthcoming subsections, as the populations
of the other clusters in the gas phase (and correspondingly on the
surface) will be statistically insignificant.

\subsection*{Cluster adsorbed onto the free-standing graphene\label{cluster_adsorb_surface}}

Next, we selected the lowest energy gas-phase cluster 6-1 from Fig.
\ref{6_atoms_charged_figure} and applied the PSO algorithm to find
the lowest energy structures this cluster can form when adsorbed onto
the graphene, when initially placed at different lateral positions
and orientations, prior to geometry relaxation.

Prior to this calculation (see the Supporting Information file), we
performed a PSO-based exploration of the best \emph{neutral} cluster
of 6 Mo atoms placed on graphene (using as its initial geometry the
one obtained in our previous work \cite{wei2023comprehensive}),
and within 1.0 eV from the best structure we found altogether 9 lowest
energy configurations with odd multiplicities between 1 and 7 (our system has an even number of electrons). Note that altogether, considering
all geometries tried by the PSO approach, more than 90 structures
were relaxed with the spread of energies of 2.24 eV. It was observed,
however, that the first five best structures have multiplicities 1 and 3 only, with the sixth structure of multiplicity 5 being by 0.4 eV higher in energy than the best structure (of multiplicity 1). Based on these preliminary calculations, when placing the best 6-1 cluster on graphene (obtained in the gas phase assuming a single positive charge, see the previous Section), 
%\ref{subsec:Lowest-energy-Mo6}), 
we limited ourselves to multiplicities 1 and 3 only. Note that the geometry of the isolated
charged cluster was used in the PSO run only when placing the cluster on
graphene  in preparing its initial geometry prior to the DFT geometry
relaxation; the latter was performed for the neutral system assuming
an immediate electron transfer from graphene, as noted previously.
%in Section \ref{subsec:General-computational-strategy}.

\begin{figure}[htbp]
\centering
\includegraphics[height=6cm]{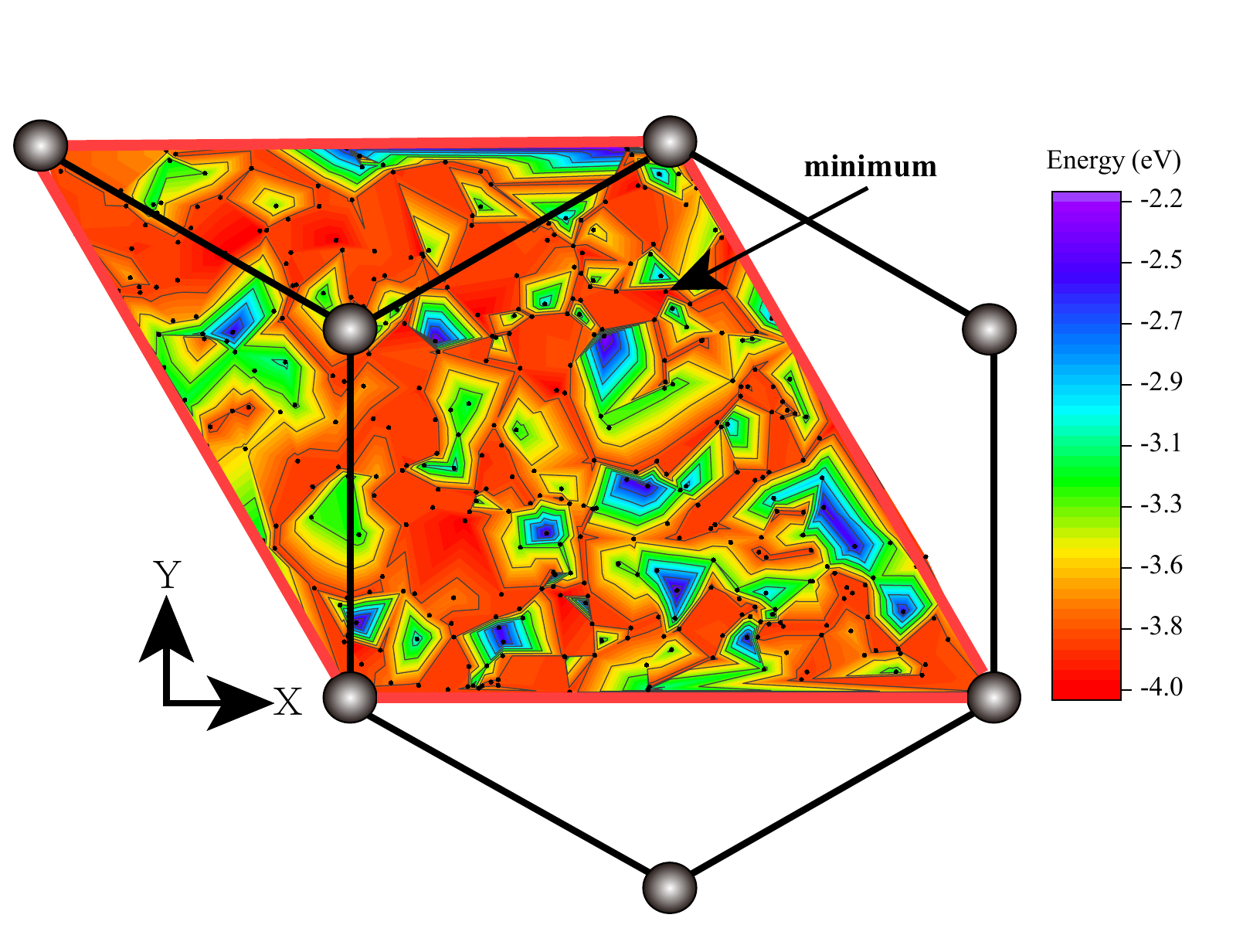}
\includegraphics[height=6cm]{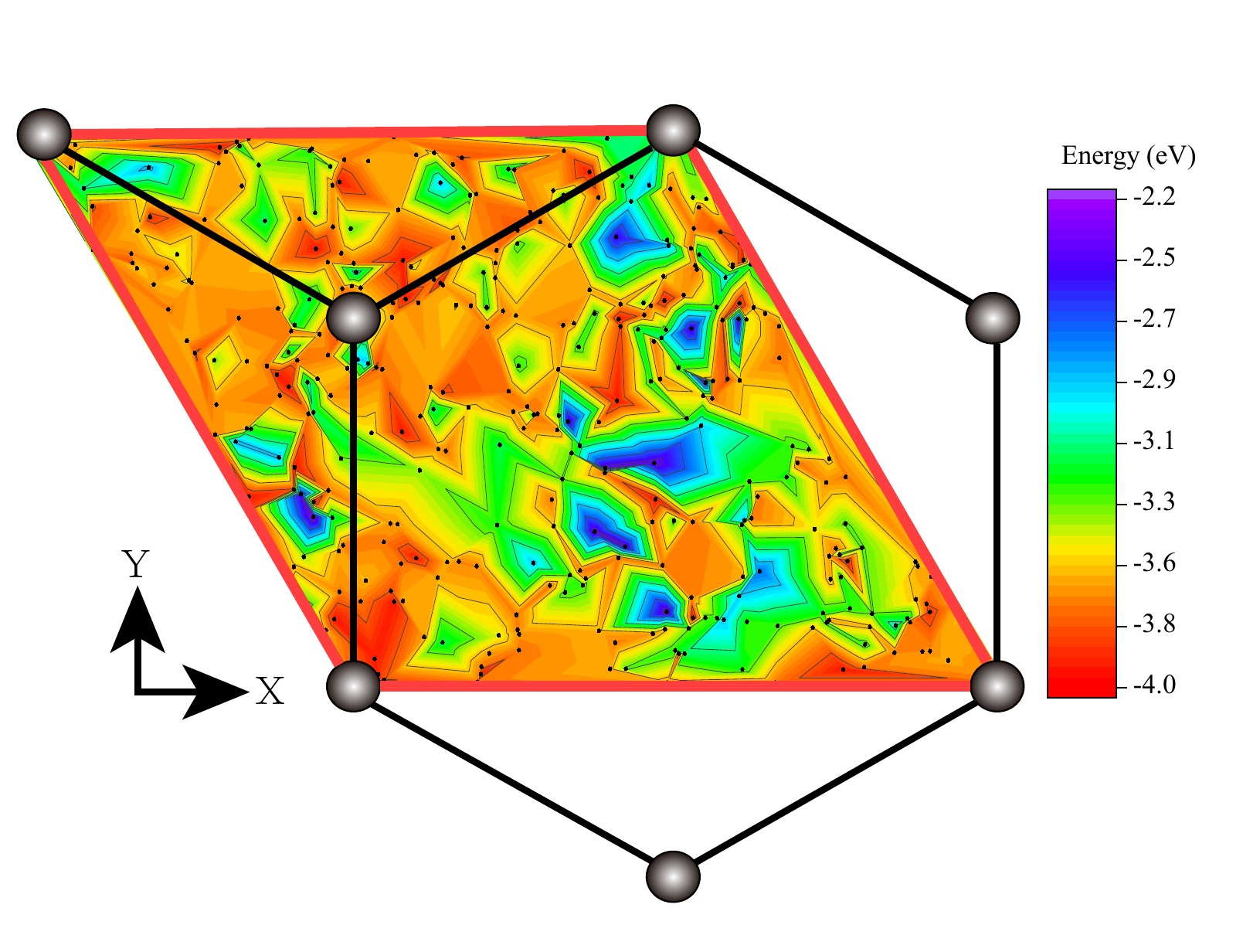}
\caption{The map of the adsorption energy of the Mo$_{6}$ cluster 6-1 adsorbed on graphene as a function of its lateral position for multiplicities $M$ being 1 (left) and 3 (right). The graphene structure (shown by connected grey circles) is superimposed on the picture for convenience.}
\label{fig:The-energy-distribution}
\end{figure}

When performing the PSO-based exploration of the potential energy
surface of our best \emph{Mo$_{6}$} cluster 6-1 on graphene (using
the geometry of the charged cluster as the initial one), 25 structures
were tried overall with a total spread of 1.8 eV in their energies.
In Fig. \ref{fig:The-energy-distribution} the distribution of these
clusters over their energies are shown as a function of the clusters'
lateral position. Note that the three angles specifying the clusters'
orientation is ignored, i.e. each point on the potential adsorption energy surface
(PES) (which is basically a projection of the whole 5-dimensional
PES onto the cluster's 2D lateral position subspace) corresponds to some values
of the angles that are most likely to be different from point to point.

\begin{figure}[htbp]
\centering
\includegraphics[width=1\textwidth]{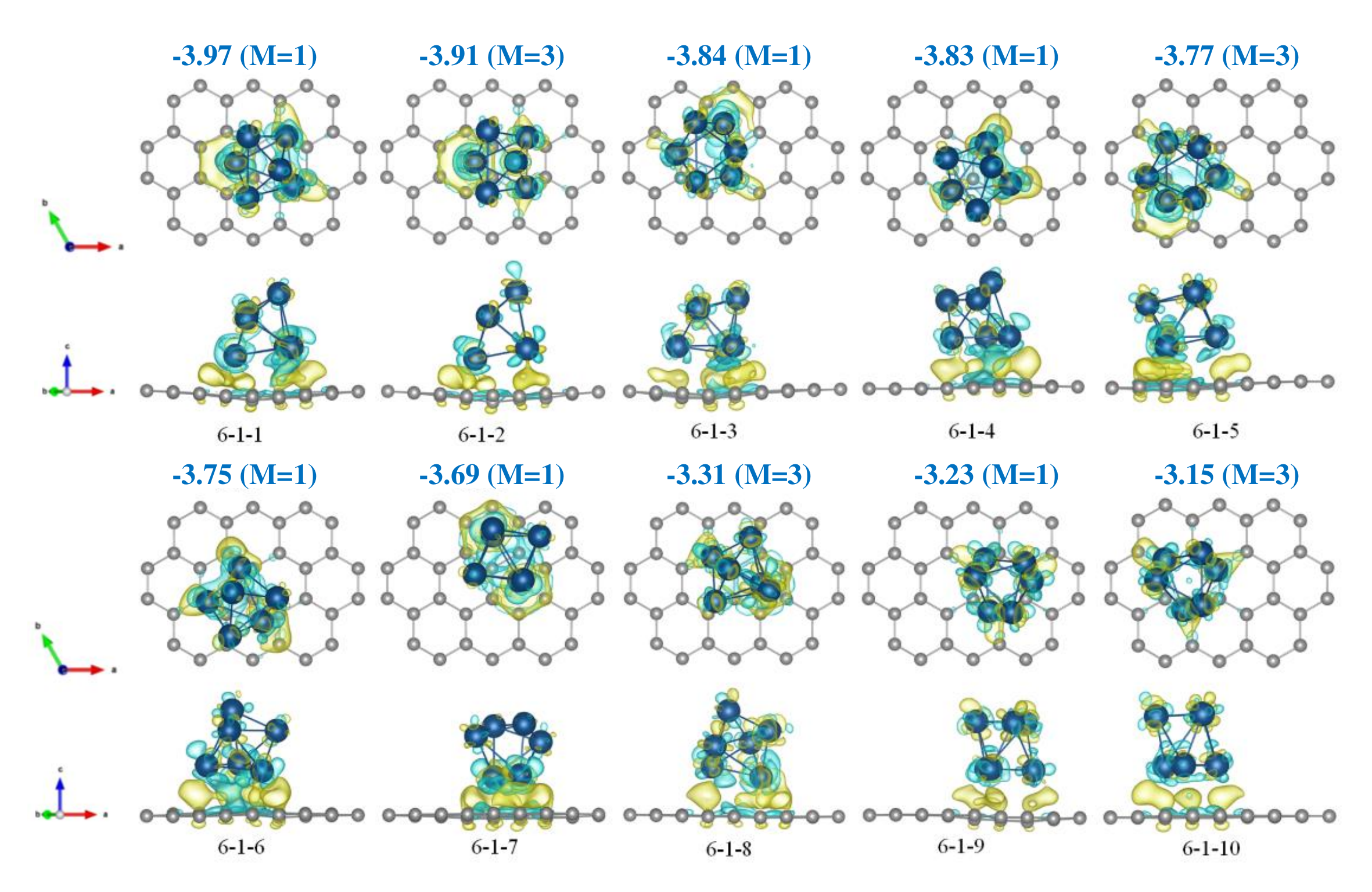}
\caption{Top (above) and side (below) views of the best ten geometries of the Mo$_{6}$ cluster adsorbed on graphene using the best 6-1 charged cluster geometry as the initial guess. The adsorption energies (in eV) and multiplicities $M$ of the clusters (in $\mu_B$) are also shown. In each case 
%Top (above) and side (below) views of 
the charge density difference is also shown.
%for the ten best Mo$_{6}$ clusters on graphene. 
The green and yellow colours correspond to depletion and excess of the density (at $\pm0.005$ Bohr$^{-3}$), respectively.}
\label{fig:charge_density_diff}
\end{figure}

The substantial variation of the PES and its complexity show, in particular,
how important it is to explore the PES in cases of complicated
systems in order to approach as close as possible the global minimum
and hence avoid non-representative results that may mislead further
study and the conclusions to be made. The ten lowest energy structures
obtained within 1 eV energy range are shown in Fig. \ref{fig:charge_density_diff}
together with their adsorption energies and multiplicities $M$. As
usual, the adsorption energy is defined as the difference 
\[
E_{ads}=E_{tot}-\left(E_{G}+E_{cl}^{0}\right)
\]
of the total energy $E_{tot}$ of the combined system (the graphene
and the cluster) and of the individually relaxed graphene $E_{G}$
and the neutral cluster $E_{cl}^{0}$; the latter energy was obtained
by considering the geometry of the relaxed charged cluster 6-1 in the neutral charge state. It
is seen that the differences in the adsorption energies among the
best four structures are relatively small, within 0.14 eV.

It is interesting to observe that for the particular system we are
considering here, the four lowest-energy structures obtained in the
PSO calculation are the same when using, as the initial geometry, either the charged
or neutral clusters  (small energy differences and hence
a different order in Figs. \ref{fig:charge_density_diff} and Fig. S1 in ESI is explained by slightly different tolerances used in the two
simulations). Note, however, that for other systems this may not be
the case.

%\begin{figure*}[htbp]
%\centering
%\includegraphics[width=1\textwidth]{graph/6_1charged_surface}
%\caption{Top (above) and side (below) views of the best ten %geometries of the Mo$_{6}$ cluster adsorbed on graphene using %the best 6-1 charged cluster geometry as the initial guess. %The adsorption energies (in eV) and multiplicities $M$ of the %clusters are also shown.}
%\label{fig:6_charged_cluster_surface}
%\end{figure*}

In Fig. \ref{fig:charge_density_diff} we also show the density difference
for selected systems. The density difference is defined in a similar
way to the definition of the adsorption energy, by subtracting from
the density of the whole system the densities of the two fragments
(the graphene and the cluster, the latter is considered as neutral);
note, however, that the geometries of the fragments correspond to
their geometry in the combined system, i.e. not individually relaxed.
It is seen that in all cases there is some charge transfer between
the clusters and graphene, as well as some redistribution of the density
within the clusters and in graphene just underneath.

%\textcolor{red}{ASB: Spin-polarized calculations reveal non-%%zero spin densities exclusively in systems with multiplicity 3 (see ESI Fig. 3). While graphene's spin density is negligible in systems 6-1-2 and 6-1-5, it's significant in the remaining %%two.}

As all our calculations were spin-polarized, we have access to the
spin densities of our systems. Only systems with multiplicity 3 demonstrated
a non-zero spin density, as expected, see Fig. S3 in ESI.
%and \textcolor{red}{ASB: shown in Fig. 3 of the ESI.} 
If in systems 6-1-2 and 6-1-5 the spin density on graphene is insignificant, in the other two systems graphene is found to be spin-polarized.
%The spin density distributions for all such systems are shown %in Fig. 
%\ref{supp-fig:6-charged-mo-on-gr-spin-density}.

\subsection*{CO molecule adsorbed on clusters on graphene}

Initially, we have considered the CO molecule adsorbing away from the cluster; however, as it
will be clear in the following, the energy of this system is almost 3 eV less favourable than when it is adsorbed on the cluster. Therefore, our next step was to adsorb the molecule within a limited lateral region with the cluster in its centre.
To this end, we used a modified version of the PSO algorithm to find the lowest
energy structures of the CO molecule at or near the 6-1 cluster adsorbed
on graphene. The algorithm in this case was modified since the rotation
of the adsorbed molecule around its axis is redundant and hence only
two angles need to be considered. Here, for simplicity, we fixed the multiplicity to
1 as above and ignored the spin polarization.

\begin{figure}[htbp]
\centering
\includegraphics[width=1\textwidth]{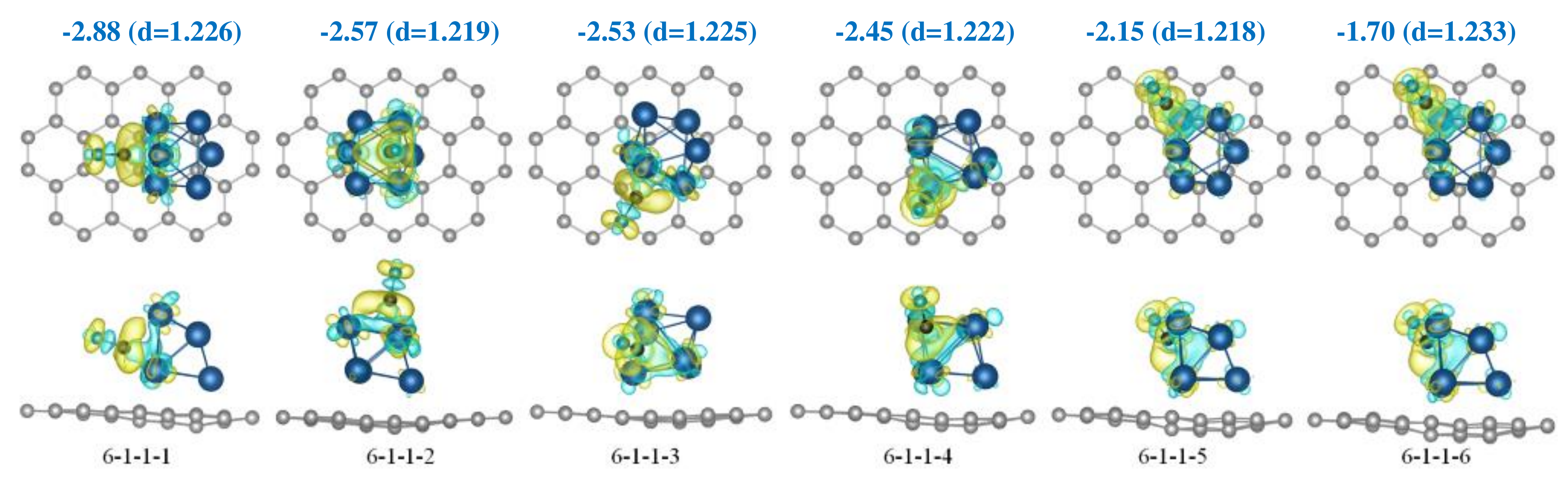}
\caption{The top (above) and side (below) views of the best six obtained geometries of the CO molecule at or near the 6-1-1 cluster on graphene. 
The adsorption energies (in eV) and the CO bond lengths $d$ (in \AA) are indicated. 
%Top (above) and side (below) views of 
The corresponding charge density difference is also shown.
%for the six best configurations of the CO molecule adsorbed on 6-1-1 %cluster that is placed on graphene. 
The green and yellow colours correspond to depletion and excess of the density (at $\pm0.005$ Bohr$^{-3}$), respectively.}
\label{fig:charge-diff-CO}
\end{figure}

As in the previous cases, more than 20 adsorption geometries of the
CO molecules on the 6-1-1 cluster on graphene were tried in our PSO
calculations, with the overall spread of their energies being of 2.67
eV. However, only six geometries within 1.2 eV from the best structure
were obtained, they are shown in Fig. \ref{fig:charge-diff-CO}. The
binding energies (defined, as before, as the energy difference between
the whole system and the two fragments, one being the cluster on graphene
and the other the individual CO molecule) and the CO bond length $d$
in each case are also shown. Note that the energy of the CO molecule adsorbed away from the cluster was found to be 2.82 eV less favourable than the best structure 6-1-1-1 with the molecule adsorbed on the cluster.

%\begin{figure*}[htbp]
%\centering
%\includegraphics[width=1\textwidth]{graph/CO_on_cluster_on_graphene}
%\caption{The top (above) and side (below) views of the best six obtained geometries of the CO molecule at or near the 6-1-1 cluster on graphene. The adsorption energies (in eV) and the CO bond lengths $d$ (in \AA) are also shown in each case.}
\label{fig:fig:charge-diff-CO}
%\end{figure*}

It is seen that the energy difference between the first (the best) and
the second configuration is 0.31 eV, which is significant rendering
the best structure 6-1-1-1 we found to be the most abundant (statistically
relevant). The energy differences between the next four configurations
are very small, making them all being almost equally probable. The
sixth geometry is by 0.45 eV less favourable than the previous geometry.
Considering the energy differences, only the best structure must be of importance 
%can be considered 
in further analysis. However, as an illustration of the possible participation of more than a single initial structure, we have also included the second-best structure 6-1-1-2 in our simulations of the reaction kinetics (see the next section). We shall see that in our particular case, its contribution can safely be neglected unless at very high temperatures (and hence of the other three structures following after it in energy). 
%, the other geometries can be safely neglected in our particular case
Of course, more structures will
have to be considered, would the energy differences with the best
structure be smaller. We also note that in all geometries the CO
molecule retains its individuality as its bond length exhibits minimal variation across different configurations.

The charge density differences between the whole system and two its
individual parts, the cluster on graphene and the CO molecule, calculated
in the geometry of the relaxed complex, for the best six geometries,
are also shown in Fig. \ref{fig:charge-diff-CO}. A considerable redistribution
of the electron density can be seen due to the bonding of the CO to the
cluster in all cases. No redistribution is found between the cluster
and graphene. The electronic density is not affected in this region
by the adsorption of the CO to the cluster, as one might expect.

\subsection*{The dissociation of the molecule}

After establishing the two most favourable structures of the CO molecule on the Mo$_{6}$ cluster adsorbed on graphene, we can consider the CO
dissociation reaction. We start by establishing a set of possible final geometries after the CO dissociation.
To this end, we ran a simple version of the PSO
algorithm that was concerned  in each case only with the lateral positions of the
O atom of the molecule near or at the cluster, i.e., we consider
only two degrees of freedom. The $z$ coordinate of the O atom
was in all cases initially taken at random between 1.6 and 2.1 $\textrm{Å}$
above the highest atom of the cluster, prior to geometry optimization. For
simplicity, the same initial position of the C atom was used in these calculations.
(Of course, in an exhaustive analysis we would  have to consider all possible
separate positions of the C and O atoms as the final states of the
dissociation reaction; this will be left for future studies.)
 In addition, we have also considered the molecule to dissociate on graphene away from the cluster. However, the energy of the final structure was found to be even less favourable (by 6.76 eV) than the energy of the CO molecule adsorbed on graphene. Hence, this mechanism for the CO dissociation can be safely disregarded.

 We shall start by considering possible reaction paths from the best 6-1-1-1 geometry. 
Overall,  as the final geometry, 12 positions of the O atom were found with the spread of
energies being 0.55 eV. Only the best two such structures will be further explored here. 

In the first structure (numbered 4), we found the O atom
being on top of the cluster with the energy by 0.67 eV lower than in the
initial state (numbered 1), and in the second structure (numbered 6) the O atom is found
at the other side of the cluster with the energy being lowered by
further 0.33 eV. This latter structure has the lowest energy we find; in it C and O atoms are positioned at the opposite sides of the cluster.

\begin{figure*}[htbp]
\centering
\includegraphics[width=1\textwidth]{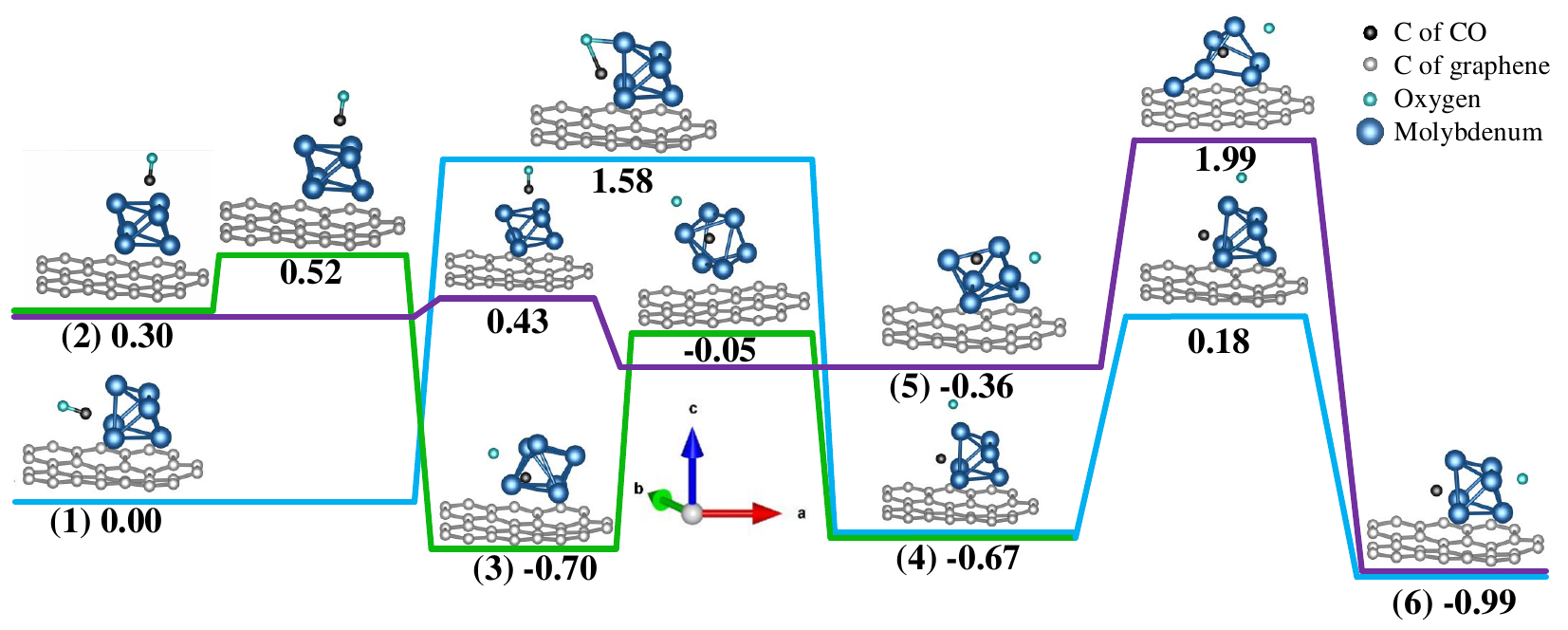}
\caption{CO molecule dissociation kinetics modelled from two different initial structures. Relative energies (in eV) of  each configuration with respect to the structure 1 are also shown. Note that different states of the system shown by horizontal lines are not drawn to the correct energy scale. All six geometries are numbered as in Eq. (\ref{rateeqs}).}
\label{fig:neb_energy}
\end{figure*}

Initially,  two NEB simulations were run, one between states 1 and 4 (3 images between the initial and final states), and another - between states  4 and  6 (4 images), see the blue curve in Fig. \ref{fig:neb_energy}.
The dissociation is governed by the first transition ($1 \rightarrow 4$) with the barrier of 1.58 eV as the second barrier for the $4 \rightarrow 6$
transition is much lower (0.85 eV). 

Next, we performed dissociation calculations from the second-best initial state 6-1-1-2 of the CO molecule on the cluster (numbered 2). 
By running an NEB simulation between states 2 and 6 we found, however, a new intermediate state (numbered 5), and therefore had to split the calculation into 
two,  $2 \rightarrow 5$  (the barrier 0.13 eV, the number of images used is 6)  and $5 \rightarrow 6$ (4 images with the barrier of 2.35 eV), see the purple curve in Fig. \ref{fig:neb_energy}.

Finally, attempts have been made to run NEB simulations between the two initial states and various intermediate minima we have found. Only one such calculation, 
namely between states 2 and 4, was successful, the green curve in Fig. \ref{fig:neb_energy}, after it was split
into two separate simulations (using by 6 images in either case) employing an additional intermediate state 3
that is lower than the best initial state by 0.7 eV.  Other calculations resulted in very large barriers (over 3 eV) due to considerable reorganisation of the cluster and both C and O atoms that was required; we do not show these simulations here. Interestingly, geometry 3 did not come out in our PSO simulations; this must be due to an insufficient number of generations used.
 % All obtained energy barriers are shown in Table \ref{tab:barriers}.

\begin{table}[h]
\centering
\begin{tabular}{|c|c|c|}
\hline 
colour in Fig. \ref{fig:neb_energy}& Transition & Barrier (eV) \tabularnewline
\hline 
\hline 
\multirow{4}{*}{Blue} & $1 \rightarrow 4$ & 1.58 \tabularnewline
\cline{2-3} 
 & $4 \rightarrow 1$ & 2.25 \tabularnewline
 \cline{2-3} 
  & $4 \rightarrow 6$ & 0.85 \tabularnewline
  \cline{2-3} 
   & $6 \rightarrow 4$ & 1.17 \tabularnewline
\hline 
\multirow{4}{*}{Green} & $2 \rightarrow 3$ & 0.22 \tabularnewline
\cline{2-3} 
 & $3 \rightarrow 2$ & 1.22 \tabularnewline
\cline{2-3} 
 & $3 \rightarrow 4$ & 0.65 \tabularnewline
\cline{2-3} 
 & $4 \rightarrow 3$ & 0.62 \tabularnewline
\hline 
\multirow{4}{*}{Purple} & $2 \rightarrow 5$ & 0.13 \tabularnewline
\cline{2-3} 
 & $5 \rightarrow 2$ & 0.79 \tabularnewline
\cline{2-3} 
 & $5 \rightarrow 6$ & 2.35 \tabularnewline
\cline{2-3} 
 & $6 \rightarrow 5$ & 2.98 \tabularnewline
\hline 
\end{tabular}
\caption{Calculated energy barriers for the specified transitions.}
\label{tab:barriers}
\end{table}

Therefore, considering  two initial geometries of the CO molecule
on the Mo$_6$ cluster, we obtained four stable final geometries of the dissociated
molecule. To determine the expected populations of the latter geometries
after the reaction (at long times), a simple rate equation analysis
can be performed.  Let $N_{1}(t)$ to  $N_{6}(t)$ be
populations of the two initial (the CO molecule is not yet dissociated) and the four final structures (C and O atoms are separated), respectively. 
%after the reaction: $N_{1}$
%corresponds to the population of states when the dissociated O atom
%is placed above the cluster and $N_{2}$ when it is placed on the
%other side of it. 
 $N_6$ corresponds to the lowest energy dissociated state of the CO molecule.
These populations can be obtained by solving the
system of linear rate equations:
\begin{equation}
\begin{cases}
\dot{N}_{1} & =-k_{1\rightarrow4}N_{1}+k_{4\rightarrow1}N_{4}\\
\dot{N}_{2} & =-k_{2\rightarrow3}N_{2}-k_{2\rightarrow5}N_{2}+k_{3\rightarrow2}N_{3}+k_{5\rightarrow2}N_{5}\\
\dot{N}_{3} & =-k_{3\rightarrow2}N_{3}-k_{3\rightarrow4}N_{3}+k_{2\rightarrow3}N_{2}+k_{4\rightarrow3}N_{4}\\
\dot{N}_{4} & =-k_{4\rightarrow1}N_{4}-k_{4\rightarrow3}N_{4}-k_{4\rightarrow6}N_{4}+k_{1\rightarrow4}N_{1}+k_{3\rightarrow4}N_{3}+k_{6\rightarrow4}N_{6}\\
\dot{N}_{5} &=-k_{5\rightarrow2}N_{5}-k_{5\rightarrow6}N_{5}+k_{2\rightarrow5}N_{2}+k_{6\rightarrow5}N_{6}\\ 
\dot{N}_{6} & =-k_{6\rightarrow4}N_{6}-k_{6\rightarrow5}N_{6}+k_{4\rightarrow6}N_{4}+k_{5\rightarrow6}N_{5}
\end{cases}
\label{rateeqs}
\end{equation}
where $k_{i\rightarrow j}=\nu\exp\left(-\beta\Delta E_{i\rightarrow j}\right)$
is the reaction rate from structure $i$ to $j$ ($=1,\dots,6$),  $\Delta E_{i\rightarrow j}$
the corresponding energy barrier deduced from Fig. \ref{fig:neb_energy} and
$\beta=1/k_BT$, where $k_B$ is the Boltzmann's constant.
All the barriers are shown in Table \ref{tab:barriers}.
 It follows from the structure of the model that, as expected, at long times  the concentrations will approach their solutions
 obtained by setting the left-hand sides of the equations to
zero and employing the conservation condition $\sum_{i=1}^6\,N_{i}(t)=1$
valid at any time $t$. 

In practical
calculations, we have used the same pre-factor $\nu=10^{13}$ s$^{-1}$, the characteristic vibrational frequency.
The pre-factors can be calculated, within the classical transition
state theory, by calculating vibrational frequencies in the minima
and the saddle points \cite{Nitzan-2014}; however, these calculations
are non-trivial for complex systems as require the precise location
of these geometries. Hence these were not attempted here as 
the exact values of the pre-factors usually have only a minor influence on
the results.

\begin{figure}[htbp]
\centering
\includegraphics[height=6cm]{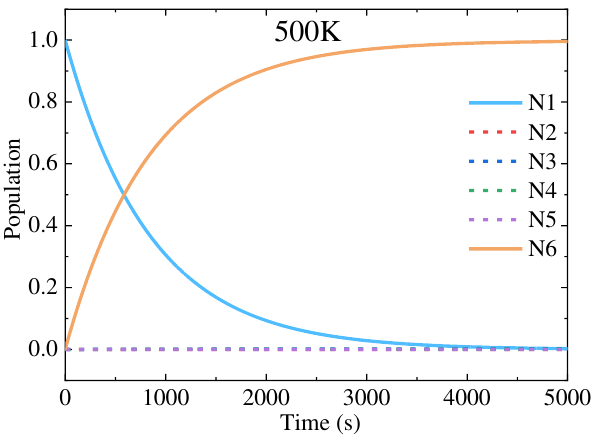}
{\includegraphics[height=6cm]{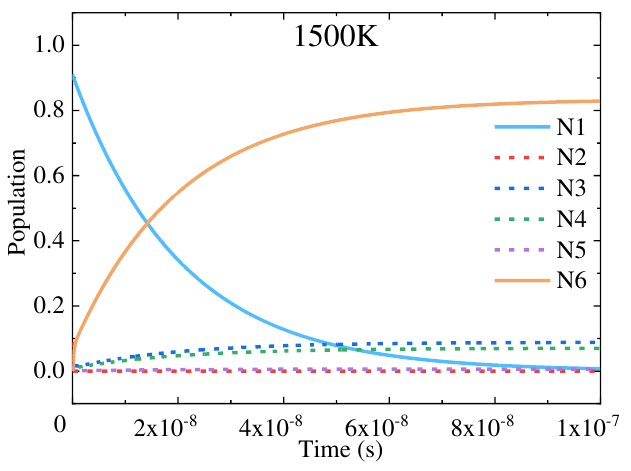}}
\caption{The time  dependence (in sec) of the populations of the CO, $N_{1}(t)$ and $N_{2}(t)$,
and of the products, $N_{i}(t)$ ($i=3,\dots,6$), during the course
of the CO dissociation reaction.}
\label{fig:populations}
\end{figure}

Initially, the populations of the intermediate and final states are set to zero.
The initial populations of the two initial states $N_1(0)$ and $N_2(0)$  are  chosen using the canonical distribution:
\begin{equation}
N_{i}(0)=Z^{-1}e^{-\beta E_{i}}\,,\,\,\,\,\,Z=\sum_{i=1}^{2}e^{-\beta E_{i}}\,,
\end{equation}
where $E_{i}$ is the binding energy of the $i$-th geometry ($i=1,2$). Clearly, only relative energies are required.
We neglected here possible kinetic transformations of the nanocluster (the so-called {\em fluxionality} effect \cite{zandkarimi2019surface,ghosh2013fluxionality,poths2024thermodynamic}) assuming its lifetime is longer than the 
characteristic reaction times (see Fig. \ref{fig:populations}).

Solving the set of equations (\ref{rateeqs}) subject to the mentioned  initial conditions, 
we obtain the time dependence of the populations as shown in Fig. \ref{fig:populations} for two temperatures, 500 K and 1500 K. 

At 500 K, we observe that practically all CO molecules dissociate, with 
the C and O atoms being at the opposite sides of the cluster, state 6. This process takes 
over an hour.  The populations of the intermediate states, as expected,
are found negligible. At 1500 K only just over 80 percent of the molecules dissociate into state 6, with some small, but noticeable, populations found 
of the intermediate states 3 and 4. The whole process at this temperature takes less than a microsecond. Similar results are obtained for 
1000 K, albeit with the reaction taking about 600 milliseconds instead (not shown).

The obtained results are to be expected due to a relatively large energy barrier (of over 1.5 eV) found in our NEB simulations for breaking the CO bond on the cluster (departure from state 1): the dissociation of CO molecules can only be observed at high enough temperatures. Still, the catalytic effect of the Mo clusters becomes evident if this barrier is compared with the CO dissociation energy of being over 11 eV \cite{PRA108}. Also, as expected, the role of the second best initial state starts playing some (still rather marginal) role only at high enough temperatures: if at 500 K the relative initial population of state 2 is only 0.1\%, at 500 K it grows to 3\% and at 1500 K to 9\%. 
These results are broadly in agreement with the observations 
\cite{CO-on-Mo-on-Al2O3} that CO molecules are not broken down by Molybdenum nano-clusters adsorbed on a thin  alumina  film unless one goes to annealing temperatures of the order of 700 K. Also, in DFT calculations \cite{Yakovkin-2009} rather large dissociation barrier was obtained for the CO dissociation on the corrugated Mo(112) surface. 
Interestingly, CO molecules were observed to dissociate, e.g.,  on the flat Mo(110) surface even at temperatures as low as 125 K \cite{CO-diss-on-Mo110}. Hence, the ability of the Mo systems to split the CO molecule is dependent on the actual Mo system, specifically, on the coordination of the Mo atoms.

\section*{Conclusion and Discussion}

In this work, we proposed a systematic global optimization search-based computational strategy for studying 
catalytic properties of sub-nanometer clusters (SNCs) on a crystal surface in nanocluster beam experiments.
The strategy consists of several well-defined stages: (i) determine the most stable clusters on the surface of interest; this step would most likely be influenced by the actual experimental procedure; note
that more than one close-in-energy structure may come out of this analysis, all need to be accounted for; (ii) for each of the possibilities found in (i) consider all possibilities for the chemical reaction
of interest catalyzed by the clusters; (iii) simulate the populations of the reactants and products after the reactions (at long times) taking into account the multitude of reaction paths from (ii).

%\textcolor{red}{ASB: While this strategy is likely known in the literature, we highlight that each stage involves numerous possibilities often overlooked in simulations. Thus, appropriate stochastic methods are crucial for proper sampling and consideration of all potential outcomes.}
Even though this strategy must have already been followed in the literature and is well-known, the message that we have tried to communicate here is that at each stage of this procedure, there are multitudes of possibilities that are rarely (if at all) considered in practical simulations, so that one has to use, at each stage, appropriate stochastic methods to achieve a proper sampling to account for all these possibilities. 
This point is highlighted by the large number of stable structures we have found with varying energies emphasizing the need for careful initial geometry selection. A "common sense" approach might lead to structures that are less relevant (e.g., of higher free energy). Note that we have not attempted to carry out free energy calculations here due to the high computational cost; only DFT total energies were computed.

%This point is highlighted by the very large number of stable structures
%we obtained at each stage, and within a wide range of energies of
%a few eV; it means that a naive approach whereby the initial geometry(ies)
%is(are) chosen using a ''common sense'' approach might lead to structures
%that are quite different from the ones that are statistically more
%relevant (of the lowest free energies). Note that we did not attempt in this
%work to actually calculate free energies; due to inevitable huge computational
%cost related to calculating vibrations for each system; only DFT total energies were computed.

To illustrate our strategy, we considered in detail the catalytic
properties of Mo$_{6}$ clusters adsorbed on the free-standing graphene
with respect to the CO molecule dissociation reaction. Specifically,
as an experimental procedure, we have examined the beam deposition
method \cite{pozzo2023interplay} in which clusters of a precise
number of atoms are evaporated on a surface from a gun without their
destruction upon impact. Note that the particular system we have chosen,
Molybdenum clusters on the free-standing graphene, may not be of significant
practical interest (even though Molybdenum is well known for its catalytic
properties \cite{H-evolution-reaction-Mo-catalysts-Rare-Met-2020,H-evolution-reaction-Mo-catalysts-Systainability-2023,Mo-watr-air-pipputants-EcoMat-2021});
it was used here on computational grounds only to illustrate the strategy.
However, the systems of real interest might be a straightforward
extension; for instance, instead of the free-standing graphene, one
may consider adsorption of Mo clusters on a graphene attached to a
transition metal surface \cite{Loi-Baraldi-2022,Loi-Baraldi-2023,Perco-Baraldi-2023}; these systems are however computationally much more demanding and
are left for future work.

Note that in performing our simulations, we have assumed a uniform distribution of the clusters in the beam. This has been done solely to simplify the computational approach. If one wishes to get beyond this approximation, it is necessary to take account of the interactions between the clusters in the beam and compute their actual trajectories \cite{PhysRevLett.71.1276}. 

Our adapted strategy for this system involved: (i) identifying the lowest-energy  (positively charged) Mo clusters using AIRSS, (ii) determining their optimal adsorption geometries on graphene using PSO, (iii) finding the lowest-energy CO adsorption sites on the most energetically favourable Mo cluster also using PSO, (iv) identifying possible final states (products) after CO dissociation near the surface employing a simplified version of PSO, (v) calculating energy barriers using NEB simulations to determine the  transition rates between the initial and final states for all minimum energy paths connecting all selected
initial and final states and (vi) performing rate equations analysis to determine the time evolution of populations of different states. For CO on a Mo$_{6}$ cluster on graphene, we found high temperatures are needed for dissociation, which broadly agrees with experiment \cite{CO-on-Mo-on-Al2O3} (although conducted for a different support). Finally, we showed that CO dissociation is unlikely to occur directly on graphene, highlighting the crucial role of Mo clusters in catalyzing this reaction.

The main message we want to deliver  is that to ensure reliable computational results in  catalysis, stochastic methods are necessary at each stage to properly sample the configurational space and identify {\em all} relevant low-energy structures. A simple initial geometry guess can lead to unrepresentative structures and misleading conclusions about catalytic properties.

% The main message from this study we believe is that to ensure reliability
% of the computational approach in studying, e.g., catalysis, as in
% the present work, stochastic methods must be used at each stage to
% properly sample the corresponding configurational phase space to determine
% the lowest energy structure(s). In many cases, there could be not one
% but several such structures with low enough energies, all of
% which need to be accounted for in the investigation (e.g., of the
% catalytic activity). For instance, a simple guess for the initial
% geometry of a cluster on a surface prior to geometry relaxation may
% lead to its geometry that is not representative (i.e., of a low population)
% and this may result in misleading conclusions in further analysis
% (e.g., of the cluster's catalytic properties).

Another point worth mentioning is that in some cases metastable configurations may be even more important when considering a given reaction, than the global minimum one.Indeed, transition rates out of some metastable states of the reactants could be higher than out of the global minimum state (the barriers lower), so that the former, not the latter, would mostly contribute to the reaction whereby exhibiting unique catalytic properties \cite{zandkarimi2019surface}. Therefore, when applying global search stochastic methods, it is essential to keep also higher energy structures and consider their role in catalysis. Two points are in order here: (i) initial populations of metastable states crucially depend on their relative energies (with respect to the global minimum state), and (ii) one may need to take account of the dynamics of reactants as well in case their lifetimes are shorter than the characteristic time of the reaction itself \cite{poths2024thermodynamic}. We have not considered the second point in this work assuming that our Mo clusters are sufficiently stable; 
however,  as an example, we inspected the first point by considering the second most favourable CO configuration on the cluster and its role in the dissociation dynamics. Even though in our case the dynamics is driven by the most favourable CO configuration, this does not need to be the case in general. 

It is appreciated that any method related to global search optimisation is computationally expensive as it is based on screening very many structures.
If a DFT based method is employed to perform total energy relaxation calculations and a modest computational resource is available, then inevitably one 
would only be limited to relatively small systems. If bigger systems are of interest, then approaches that make the calculations more efficient must be sought.
In particular, methods based on Machine Learning \cite{ma2020machine,toyao2019machine} or relying on force fields of {\em{ab initio}} quality \cite{deringer2020general} must be considered as the most important practical ingredient of the proposed approach.
Importantly, however, the general computational strategy outlined in this work would remain the same.

We hope that this study will be of utility to a wide class of researchers
in computational surface physics and chemistry.

\begin{acknowledgement}

Y.W. is grateful for the funding from the China Scholarship Council.
The calculation resource is supported via our membership of the UK's
HEC Materials Chemistry Consortium, which is funded by EPSRC (EP/R029431
and EP/X035859), this work used HPC of ARCHER2 and YOUNG. ARCHER2
UK National Supercomputing Service (http://www.archer2.ac.uk), and
the UK Materials and Molecular Modeling Hub for computational resources,
MMM Hub, which is partially funded by EPSRC (EP/T022213 and EP/W032260).
We would also like to thank Alessandro Baraldi for useful discussions.
\end{acknowledgement}

\section{Supporting Information}
Generating cluster geometry on graphene in PSO method; Adsorption results based on the best neutral Mo$_{6}$ cluster; Spin density results based on the best charged Mo$_{6}$ cluster; The full path for NEB calculation.

\bibliography{achemso-demo}

\end{document}

\begin{textblock}{10}(1,1)
    \LARGE\textbf{Supporting Information}
\end{textblock}

\title{Global Optimization of Molybdenum Subnanoclusters on Graphene: a Consistent Approach Towards Catalytic Applications}
\author{Yao Wei}
\affiliation{Theory and Simulation of Condensed Matter (TSCM), King's College London,
Strand, London WC2R 2LS, United Kingdom}
\author{Alejandro Santana-Bonilla}
\affiliation{Theory and Simulation of Condensed Matter (TSCM), King's College London,
Strand, London WC2R 2LS, United Kingdom}
\author{Lev Kantorovich}
\email{lev.kantorovitch@kcl.ac.uk}

\affiliation{Theory and Simulation of Condensed Matter (TSCM), King's College London,
Strand, London WC2R 2LS, United Kingdom}
\maketitle

\section{Generating cluster geometry on graphene in PSO method}

This work establishes a direct method for deriving the atomic positions of Mo clusters on graphene utilizing the five-dimensional search space inherent in the Particle Swarm Optimization (PSO) algorithm. The degrees of freedom can be split into 2 lateral positions of the cluster's centre of mass ($X_{cm}$
and $Y_{cm}$) and 3 Euler's angles ($u$, $v$ and $w$) represent the possible rotations in the space (see Table
\ref{tab:Euler-angles}). The vertical ($Z_{cm}$) position of the
cluster's centre of mass is defined such that the distance between
the unrelaxed graphene sheet and the lowest atom of the cluster (for the given orientation) is 3.0 Å. This is sufficient for
the cluster to experience a noticeable vertical force from graphene.

\begin{table}[H]
\centering{}\caption{Euler angles. \label{tab:Euler-angles}}
\begin{tabular}{|c|c|c|}
\hline 
\textbf{Axis of rotation} & \textbf{Euler angle name} & \textbf{Euler angle symbol}\tabularnewline
\hline 
$x$ & Roll & \textbf{$u$}\tabularnewline
\hline 
$y$ & Pitch & $v$\tabularnewline
\hline 
$z$ & Yaw & $w$\tabularnewline
\hline 
\end{tabular}
\end{table}

If $\mathbf{R}_{0}=(x_{0},y_{0},z_{0})$ are Cartesian coordinates
of an atom of the cluster in the laboratory coordinate system in which the cluster is defined in the gas phase, with its centre of mass being at the zero point, then the coordinates $\mathbf{R}=(x,y,z)$ of that atom after rotation by the three Euler angles and translation by the vector $\mathbf{R}_{cm}=(X_{cm},Y_{cm},Z_{cm})$ are found as 
\[
\mathbf{R}=\mathbf{R}_{cm}+\mathbf{U}_{z}(w)\mathbf{U}_{y}(v)\mathbf{U}_{x}(u)\mathbf{R}_{0}\,,
\] where the three rotations around the $x$, $y$ and $z$ axes are
defined by their respective matrices as follows:

\begin{equation}
\begin{aligned}\mathbf{U}_{x}(u) & =\begin{bmatrix}1 & 0 & 0\\
0 & \cos(u) & -\sin(u)\\
0 & \sin(u) & \cos(u)
\end{bmatrix}\\
\mathbf{U}_{y}(v) & =\begin{bmatrix}\cos(v) & 0 & \sin(v)\\
0 & 1 & 0\\
-\sin(v) & 0 & \cos(v)
\end{bmatrix}\\
\mathbf{U}_{z}(w) & =\begin{bmatrix}\cos(w) & -\sin(w) & 0\\
\sin(w) & \cos(w) & 0\\
0 & 0 & 1
\end{bmatrix}\,.
\end{aligned}
\end{equation}

The yaw and roll angles are in the range - $\pi$ to + $\pi$ , while the pitch angle varies between -$\pi$/2 and +$\pi$/2.
The displacement vector $\mathbf{R}_{cm}$ is defined via the fractional coordinates associated with the surface unit cell, with two fractional coordinates varied between 0 and 1.

\section{Adsorption results based on the best neutral Mo$_{6}$ cluster}

\begin{figure*}[htbp]
\begin{centering}
\includegraphics[width=15cm]{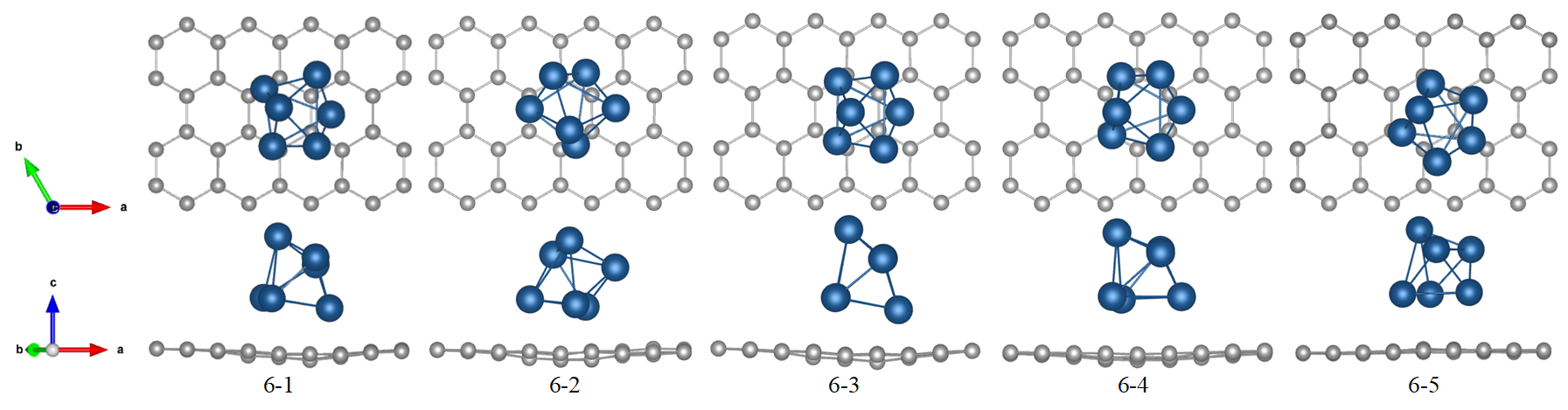}
\par\end{centering}
\begin{centering}
\includegraphics[width=15cm]{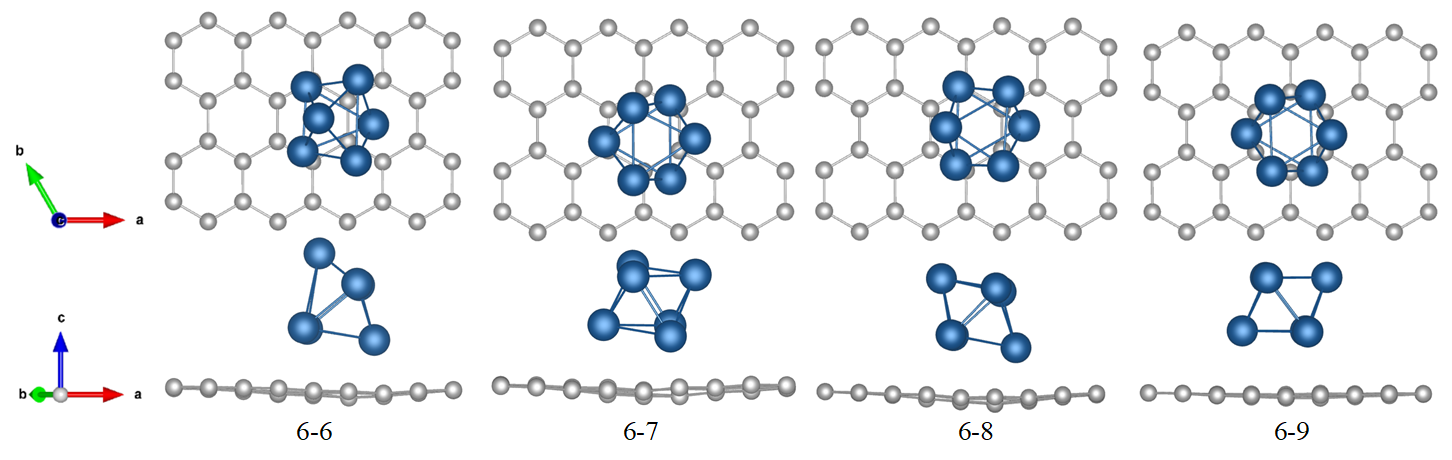}
\par\end{centering}
\centering{}
\caption{Minimum energy systems (1-9) of 6 Mo atoms adsorbed on graphene. 
\label{fig:6-01_atoms_on_surface}}
\end{figure*}

\begin{figure*}[h]
\centering{}\includegraphics[width=15cm]{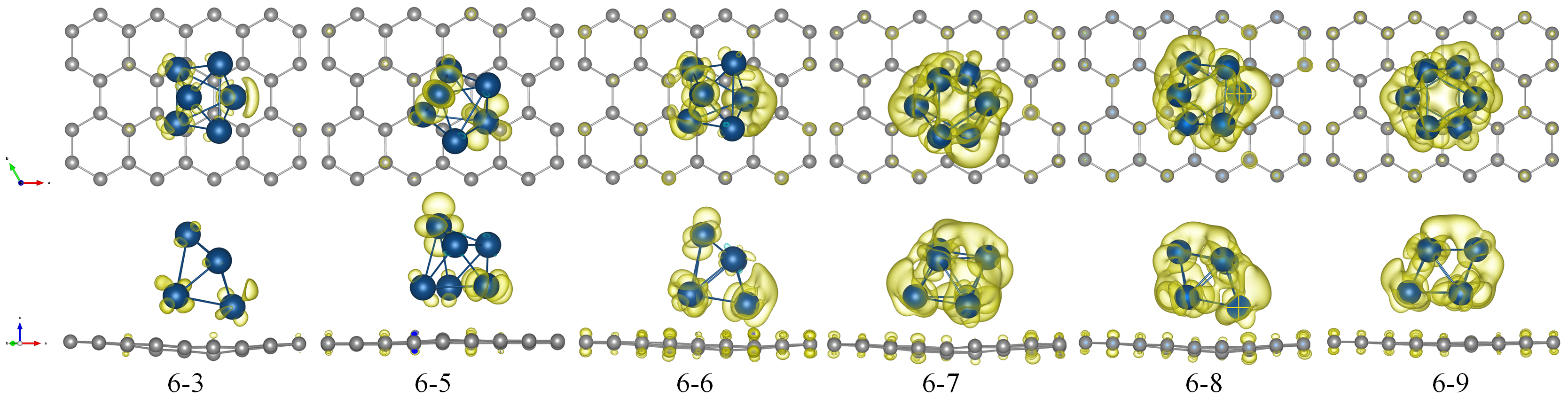} 
\caption{The spin densities of 6-atom Mo clusters adsorbed on graphene. \label{fig:6-01-spin}}
\end{figure*}

Figure \ref{fig:6-01_atoms_on_surface} presents the nine most energetically favorable atomic configurations (within 1.0 eV of the absolute minimum) for the Mo$_{6}$ cluster adsorbed on graphene and whose initial geometry 
corresponded to that in the gas phase assuming its neutrality. The energies and the multiplicities of the systems are shown in Table \ref{tab:6atoms-graphene-energy_1}. The spin densities of the relevant clusters are shown in Fig. \ref{fig:6-01-spin}.

\begin{table*}[h]
\centering{}\caption{The adsorption energy $E$ (in eV) and spin multiplicity $M$ of 9
lowest energy structures of Mo$_{6}$ clusters on graphene found.
\label{tab:6atoms-graphene-energy_1}}
\begin{tabular}{c|ccccccccc}
\hline 
 & 6-1 & 6-2 & 6-3 & 6-4 & 6-5 & 6-6 & 6-7 & 6-8 & 6-9\tabularnewline
\hline 
adsorption energy (eV) & 3.77 & 3.75 & 3.70 & 3.59 & 3.45 & 3.37 & 2.97 & 2.97 & 2.87\tabularnewline
\hline 
Spin multiplicity & 1 & 1 & 3 & 1 & 3 & 5 & 7 & 7 & 7\tabularnewline
\hline 
\end{tabular}
\end{table*}

\clearpage{}
\section{Spin density results based on the best charged Mo$_{6}$ cluster}

\begin{figure}[htbp]
\centering
\includegraphics[width=1\textwidth]{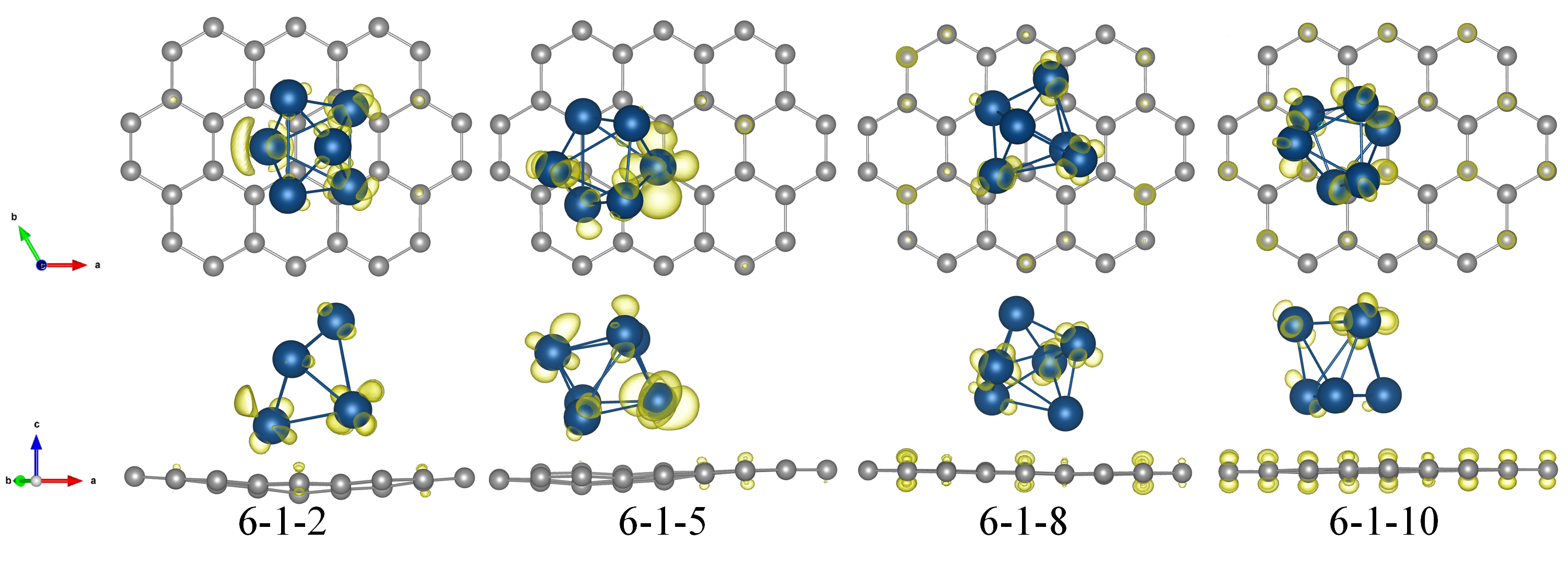}
\caption{Top (above) and side (below) views of the spin density (at $\pm0.005$ Bohr$^{-3}$) for selected Mo$_{6}$ clusters on graphene that demonstrate non-zero spin density ($M=3$).}
\label{spin-density}
\end{figure}

The spin density for selected clusters on graphene obtained assuming, as the cluster's initial geometry, its geometry with the single positive charge, is shown in Fig. \ref{spin-density}.

\clearpage{}

\section{The full path for NEB calculation}

\begin{figure}[H]
\begin{centering}
\includegraphics[width=12cm]{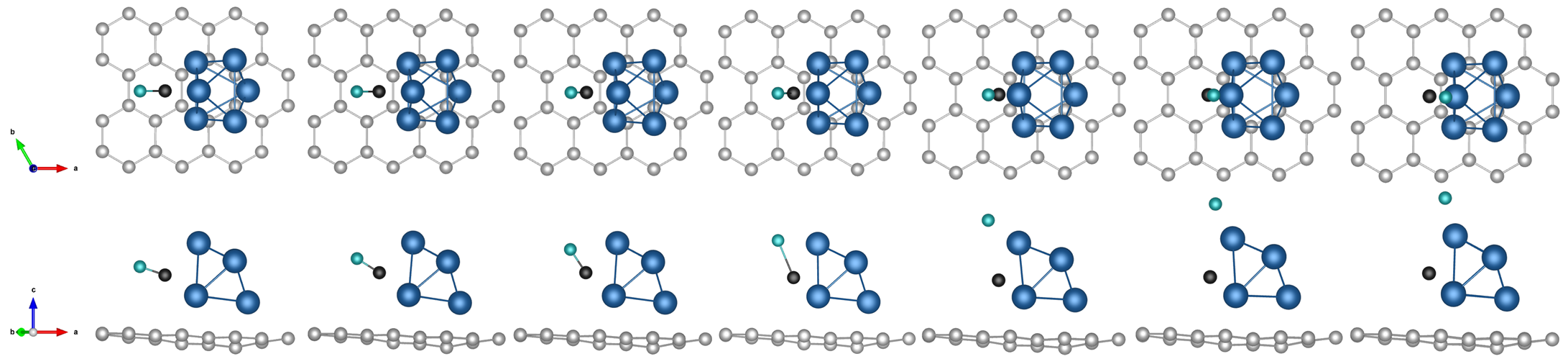}
\par\end{centering}
\caption{All geometries along the first minimum energy path corresponding to the dissociation of the CO molecule on the Mo$_{6}$ cluster on graphene with the O atom (green) placed on top of the cluster in the final state. \label{fig:neb1}}
\end{figure}

\begin{figure}[H]
\begin{centering}
\includegraphics[width=12cm]{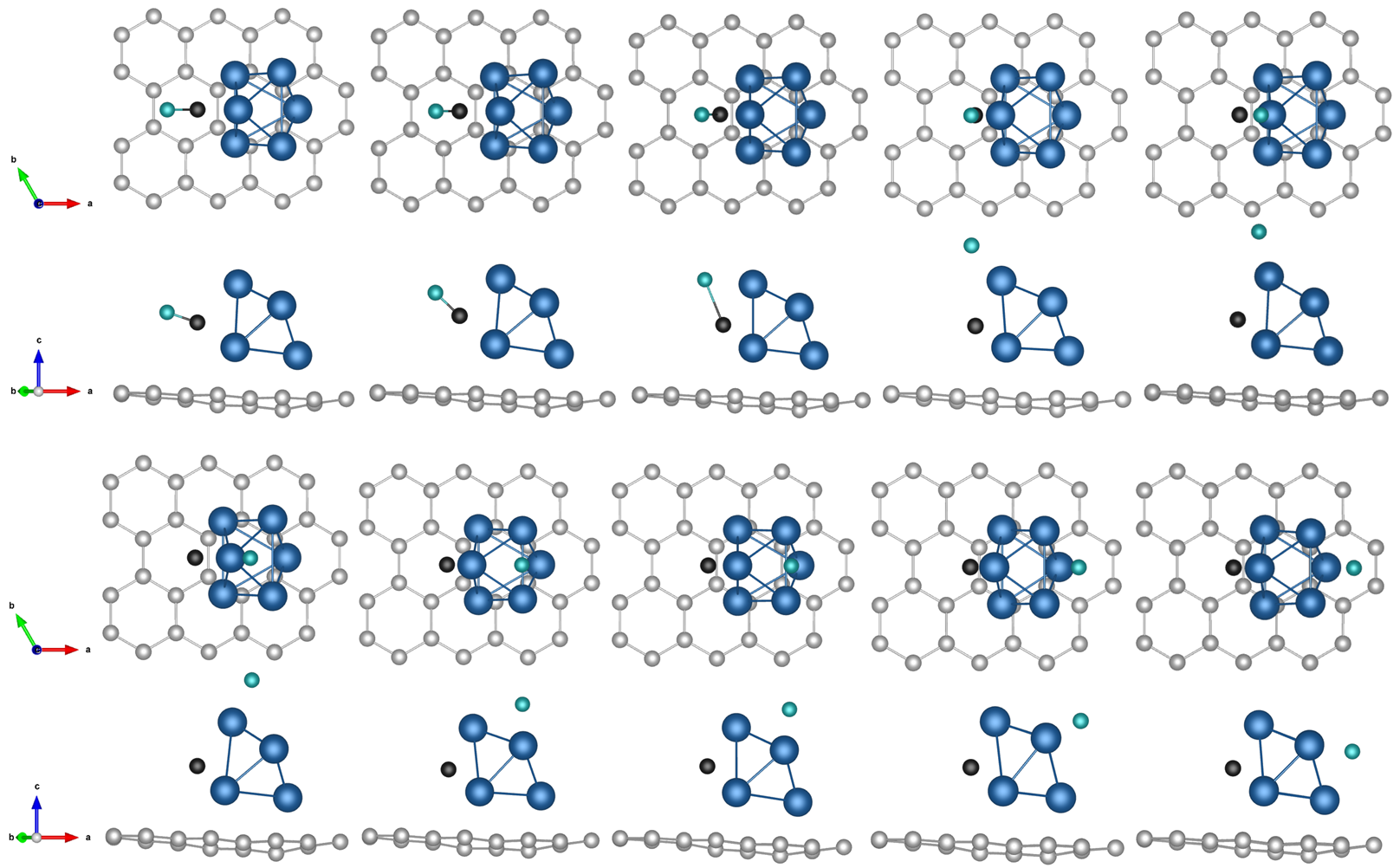}
\par\end{centering}
\caption{All geometries along the second minimum energy path corresponding
to the dissociation of the CO molecule on the Mo$_{6}$ cluster on
graphene with the O atom (green) placed on the other side of the cluster.\label{fig:neb2}}
\end{figure}

In Figures \ref{fig:neb1} and \ref{fig:neb2}, we display the different geometries employed in the Nudge Elastic Band (NEB) calculations. Figure S4 shows the seven images 
used to compute the path, while Figure S5 displays
the ten images computed for the second case as discussed in section \textbf{CO molecule absorbed on clusters on graphene}.

